\documentclass[11pt]{article}
\usepackage{graphicx}

\begin{document}

\title{Lessons from failures to achieve what was possible in the twentieth century physics}
\author{Vesselin Petkov \\
Science College, Concordia University\\
1455 De Maisonneuve Boulevard West\\
Montreal, Quebec, Canada H3G 1M8\\
E-mail: vpetkov@alcor.concordia.ca}
\date{}
\maketitle


\begin{abstract}
For several decades there has been no breakthrough in fundamental physics as revolutionary as relativity and
quantum physics despite the amazing advancement of applied physics and technology. By discussing several
examples of what physics could have achieved by now, but failed, I will argue that the present state of
fundamental physics is not caused by the lack of talented physicists, but rather by problematic general views on
how one should do physics. Although it appears to be widely believed that such general views cannot affect the
advancement of physics I would like to draw the attention of the younger generation of physicists to three
reasons that might have been responsible for failures in the past and might cause problems in the future: (i)
misconceptions on the nature of physical theories, (ii) underestimation of the role of conceptual analyses so
successfully employed by Galileo and Einstein, and (iii) overestimation of the predictive power of mathematics in
physics.
\end{abstract}

In recent years physicists have started to express publicly their concerns that not everything has been quite fine
with the advancement of physics in the last several decades \cite{smolin}. Although not all agree with such
concerns I think it is an undeniable fact that in this period there has been no revolutionary breakthrough in
fundamental physics of the type of relativity and quantum physics. A convenient answer might be that more
experimental evidence is required to make a breakthrough possible. But what if it turns out that all conditions
necessary for a new physical theory have already been present but physicists have been failing to process
successfully the existing theoretical and experimental evidence? A concise analysis of several missed
opportunities in the twentieth century physics due to what I think are inadequate views on doing physics seems
to suggest that this possibility should be taken seriously.

Physicists generally agree that physics describes and tries to explain the world. However, this appears to be the
only consensus. Physicists' views on the nature of physical theories vary widely but I think the views which
might prevent physicists who hold them from arriving at new results are those that regard our theories only as
good descriptions of the world and that insist on not considering the ``most successful abstractions to be real
properties of our world'' \cite{mermin} (one can even come across extreme views that physics tells nothing
about the \textit{existence} of what it describes). I think these are unproductive views negatively affecting not
only the advancement of physics, but also the physicists who share them and who often pay a high personal
price. Here are two examples from the history of the twentieth century physics.

It was precisely the view, that successful abstractions should not be regarded as representing something real,
that prevented Lorentz from discovering special relativity. He believed that the time $t$ of an observer at rest
with respect to the aether (which is a genuine example of reifying an unsuccessful abstraction) was the true
time, whereas the quantity $t'$ of another observer, moving with respect to the first, was merely an abstraction
that did not represent anything real in the world. Lorentz himself admitted the failure of his approach
\cite{lorentz}:

\begin{quote}
The chief cause of my failure was my clinging to the idea that the variable $t$ only can be considered as the
true time and that my local time $t'$ must be regarded as no more than an auxiliary mathematical quantity. In
Einstein's theory, on the contrary, $t'$ plays the same part as $t$; if we want to describe phenomena in terms
of $x', y', z', t'$ we must work with these variables exactly as we could do with $x, y, z, t$.
\end{quote}

The second example is especially relevant now since this year we celebrate the one hundredth anniversary of
the publication of Minkowski's paper ``Space and time'' in 1909 \cite{minkowski}. A century after Minkowski,
for some physicists spacetime is nothing more than a four-dimensional mathematical space which does not
represent a real four-dimensional world. In a recent \textit{Reference Frame} N. David Mermin wrote that
``spacetime is an abstract four-dimensional mathematical continuum'' \cite{mermin} and pointed out that
regarding spacetime only as an abstract concept was a good example of not ``inappropriately reifying our
successful abstractions'' \cite{mermin}. However, if this were the case and spacetime were merely a
mathematical space with no counterpart in the world, we would celebrate Poincar\'{e}, not Minkowski, because
it was Poincar\'{e} who first noticed (before July 1905) that the Lorentz transformations had a geometric
interpretation as rotations in what he seemed to have regarded as an \textit{abstract} four-dimensional space
\cite[p. 168]{poincare}.

The reason of why Poincar\'{e} failed to develop that idea seems to be even more complicated than that in the
case of Lorentz. Poincar\'{e} appeared to have seen nothing revolutionary in the idea of a mathematical
four-dimensional space since he believed that our physical theories are only convenient descriptions of the world
and therefore it is really a matter of \textit{convenience} which theory we would use in a given situation.
Poincar\'{e}'s failure to comprehend the profound physical meaning of the principle of relativity and of the
geometric interpretation of the Lorentz transformations is one of the examples in the history of physics when an
inadequate philosophical position prevents a scientist (even as great as Poincar\'{e}) from making a discovery.

This should be specifically stressed because physicists often think that they do not need any philosophical
position for their research. In order to make sure that inadequate philosophical views do not prevent us
(especially young physicists) from arriving at important new results I think it would be helpful if from time to
time we recall similar cases from the history of science or at least what the philosopher Daniel Dennett said on
this issue: ``Scientists sometimes deceive themselves into thinking that philosophical ideas are only, at best,
decorations or parasitic commentaries on the hard, objective triumphs of science, and that they themselves are
immune to the confusions that philosophers devote their lives to dissolving. But there is no such thing as
philosophy-free science; there is only science whose philosophical baggage is taken on board without
examination'' \cite{dennett}.

Had Minkowski believed, like Poincar\'{e}, that uniting space and time into a
four-dimensional space was only a convenient mathematical abstraction, he
would not have written a paper whose title and content were devoted to
something the main idea of which had already been published by Poincar\'{e}
two years (and probably written three years) before Minkowski's talk on space
and time given in September 1908, and would not have begun his paper with
the now famous introduction, which unequivocally announced the revolution
in our views on space and time: ``Henceforth space by itself, and time by
itself, are doomed to fade away into mere shadows, and only a kind of union
of the two will preserve an independent reality'' \cite[p. xv]{minkowski2}.

Unlike Poincar\'{e}, Minkowski appears to have realized that special relativity, particularly relativity of
simultaneity (which implies the existence of \textit{many} spaces), would be \textit{impossible} in a
three-dimensional world. \textit{As a three-dimensional space constitutes a single class of absolutely
simultaneous events, if the world were three-dimensional (i.e., if spacetime were not representing a real
four-dimensional world and were just an abstract mathematical space), there would exist one absolute space
and one absolute class of simultaneous events in contradiction with relativity}. If one tends to think that
Minkowski had not done more than Poincar\'{e} on the unification of space and time and that relativity does not
force us to regard the world as four-dimensional, it would be refreshingly helpful to examine the argument in
the italicized sentence (which appears to have made Minkowski realize the profound physical meaning of
relativity of simultaneity) and to try to demonstrate that the theory of relativity is possible in a
three-dimensional world.

When we take into account that relativity of simultaneity does imply that,
having different classes of simultaneous events, observers in relative motion
have \textit{different spaces}, we can better understand why Minkowski noted
``neither Einstein nor Lorentz made any attack on the concept of space''
\cite[p. xxv]{minkowski2} and emphasized that ``We would then have in the
world no longer \emph{the} space, but an infinite number of spaces,
analogously as there are in three-dimensional space an infinite number of
planes. Three-dimensional geometry becomes a chapter in four-dimensional
physics. You see why I said at the outset that space and time are to fade
away into shadows, and that only a world in itself will subsist'' \cite[pp.
xxi]{minkowski2}.

A hundred years ago Minkowski gave us the correct relativistic picture of the
world where the geodesic worldtubes of what we perceive as freely moving
three-dimensional macroscopic bodies are in reality four-dimensional
objects, not ``abstract geometric constructions'' \cite{mermin}. That
macroscopic physical bodies are real four-dimensional worldtubes is the
essence of Minkowski's explanation of the deep physical meaning of length
contraction -- as shown in Fig. 1 of his paper (the right-hand part of which is
reproduced in Fig. 1 here) length contraction would be \textit{impossible} if
the worldtubes of the two Lorentzian electrons, represented by the vertical
and the inclined bands in Fig. 1, did not exist and were nothing more than
abstract graphical constructions. To see this more clearly consider only the
electron represented by the vertical worldtube. The three-dimensional
cross-section $PP$, resulting from the intersection of the electron's worldtube
and the space of an observer at rest with respect to the electron, is the
electron's proper length. The three-dimensional cross-section
$P^{\prime}P^{\prime}$, resulting from the intersection of the electron's
worldtube and the space of an observer moving with respect to the electron,
is the relativistically contracted length of the electron measured by that
observer. Minkowski stressed that ``this is the meaning of Lorentz's
hypothesis of the contraction of electrons in motion'' \cite[p. xxv]{minkowski2}
and ``that the Lorentzian hypothesis is completely equivalent to the new
conception of space and time, which, indeed, makes the hypothesis much
more intelligible'' \cite[p. xxiii]{minkowski2}. The worldtube of the electron
must be real in order that length contraction be possible because, while
measuring the same electron, the two observers in relative motion measure
\textit{two} three-dimensional electrons represented by the cross-sections
$PP$ and $P^{\prime}P^{\prime}$ in Fig. 1.

\begin{figure}[h]
\centering
\includegraphics[height=4.5cm]{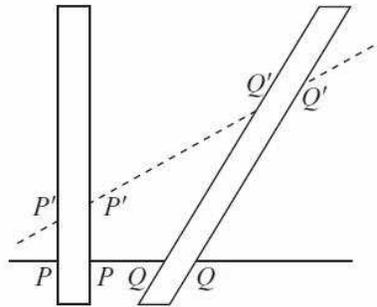}
\caption{In Minkowski's paper ``Space and Time'' Fig. 1, part of which is reproduced here, represents two
Lorentzian electrons by their worldtubes or as Minkowski called them (world) bands.}\label{mink1}
\end{figure}

This is not so surprising when one takes into account relativity of simultaneity and the fact that an extended
three-dimensional object is defined in terms of \textit{simultaneity} -- all parts of the electron taken
simultaneously at a given moment. If the worldtube of the electron were indeed an abstract geometric
construction and what existed was a single three-dimensional electron (a single class of simultaneous events)
represented by the proper cross-section $PP$, both observers would measure the same three-dimensional
electron, i.e. the same class of simultaneous events in contradiction with relativity.

\textbf{1. Missed opportunities due to failing to resolve the issue of the dimensionality of the world according to
relativity.} One of the main failures of the twentieth century physics is not taking seriously Minkowski's
arguments for the reality of spacetime or as he called it perhaps more appropriately ``the absolute world.'' For
some physicists this is not a failure at all since for them the question of whether or not spacetime is an abstract
concept or represents a real four-dimensional world is not a physical question since it rather belongs to
philosophy. Such a view is worrisome for several reasons. First, I wonder how many physicists would subscribe
to this view if directly asked whether the dimensionality of the world is a philosophical question (especially now
when physicists freely talk about multidimensional spaces and the multiverse). Second, I think such a view
deprives physicists from arriving at a deeper understanding of what relativity is telling us about the world. Third,
this particular view has been responsible for not arriving at some important results as I intend to show here.

\textit{1.1. Dimensionality of the world and inertia.} Inertia has been an open question for centuries and a
golden opportunity to finally start resolving it in the twentieth century has been omitted because of an
inadequate (and therefore unproductive) view on the nature of physical theories, namely the view that
spacetime and worldtubes are theoretical constructions that do not represent anything real.

If Minkowski's explanation of length contraction is carefully examined, understood, and adopted, it becomes
clear that this relativistic effect would be impossible if the worldtube of the contracted body did not exist. Then,
when the reality of the worldtube of an accelerating body is taken into account, one can easily link two facts: (i)
the body resists its acceleration, and (ii) the worldtube of the accelerating body is deformed (not geodesic). As
a real four-dimensional worldtube should resist its (static) deformation (like any deformed three-dimensional
rod) it is natural to ask whether inertia originates from the four-dimensional stress arising in the statically
deformed worldtube.

Calculations (i) in terms of the four-dimensional stress tensor, (ii) in the case of the classical electron, and (iii) a
semi-classical calculation in quantum field theory, all appear to show that the static restoring force ``arising'' in
the deformed worldtube of an accelerating body has the form of the inertial force \cite[Chap. 9]{petkov}. In the
case of a body at rest on the Earth's surface its deformed worldtube gives rise to a static restoring force that is
precisely $\textbf{F}=m\textbf{g}$. So taking into consideration the reality of worldtubes of macroscopic
physical bodies implies that the equivalence of inertial and gravitational forces and masses may be explained by
four-dimensional stresses arising in the deformed worldtubes of non-inertial bodies (i.e. in the worldtubes that
are deviated from their geodesic shape). The worldtube of a body falling in a gravitational field is geodesic
(\textit{not} deformed) and the body does not resist its fall.

Minkowski would be thrilled to examine the possibility that inertia may be another manifestation of the
four-dimensionality of the world, in addition to the kinematic relativistic effects, which he believed were such
manifestations.

In fact, I think there were two other failures to initiate the assault on the origin of inertia in the twentieth
century. Those missed opportunities had been caused, in my view, mostly due to underestimation of the role of
conceptual analyses in physics. Since regarding inertia as originating from a four-dimensional stress that arises
in the deformed worldtubes of non-inertial macroscopic bodies provides only a phenomenological picture of what
causes a body to resist its acceleration, the next question will be to reveal what gives rise to the
four-dimensional stress. As the three-dimensional stress in a deformed three-dimensional rod is caused by
electromagnetic forces acting on the atoms of the deformed rod that are deviated from their equilibrium
positions, it is natural to search for a similar mechanism in the case of the four-dimensional stress.

The first attempt to explain the origin of inertia, that could be regarded as a candidate for such a mechanism,
started on its own long before Minkowski. In 1881 Thomson \cite{thomson} first realized that a charged particle
was more resistant to being accelerated than an otherwise identical neutral particle and conjectured that inertia
can be reduced to electromagnetism. This conjecture had been developed in the framework of the classical
electron theory into what is now known as the classical electromagnetic mass theory of the electron (for details
see \cite{rohrlich90,jammer2000,butler}). In this theory inertia is regarded as originating from the interaction of
the electron with its own electromagnetic field.

The electromagnetic mass theory of inertia correctly predicts the experimental fact that at least part of the
inertia and inertial mass of every charged particle is electromagnetic in origin. As Feynman put it: ``There is
definite experimental evidence of the existence of electromagnetic inertia\ -- there is evidence that some of the
mass of charged particles is electromagnetic in origin'' \cite{feynman}. And despite the fact that, at the
beginning of the twentieth century, many physicists recognized ``the tremendous importance which the concept
of electromagnetic mass possesses for all of physics'' \cite{fermi22a}, it has been inexplicably abandoned after
the advent of relativity and quantum mechanics. And that happened even though the classical electron theory
predicted that the electromagnetic mass increases with increasing velocity \textit{before\/} the theory of
relativity, yielding the correct velocity dependence, and that the relation between energy and mass is
$E=mc^{2}$ \cite[pp. 28-3, 28-4]{feynman} (in fact, the relationship between mass and energy contained a
factor of 4/3, which prompted Feynman to write ''It is therefore impossible to get all the mass to be
electromagnetic in the way we hoped. It is not a legal theory if we have nothing but electrodynamics'' \cite[p.
28-4]{feynman}; but he was unaware that the factor of 4/3 had already been accounted for \cite {rohrlich90}).

In my view, there had been no justification for abandoning the classical electron theory without a thorough
analysis of how it made the predictions mentioned in the above paragraph and especially its major prediction of
the electromagnetic mass of charged particles. A completely wrong theory cannot make so many correct
predictions. Today ``the state of the classical electron theory reminds one of a house under construction that
was abandoned by its workmen upon receiving news of an approaching plague. The plague in this case, of
course, was quantum theory. As a result, classical electron theory stands with many interesting unsolved or
partially solved problems'' \cite[p. 213]{pearle}.

The other failure to start dealing with the open question of inertia is caused, I think, again by underestimation
of the role of conceptual analyses in physics. According to the Standard Model the fundamental interactions are
mediated by the exchange of virtual quanta. However, a rigorous conceptual analysis of how exactly the
exchange of virtual quanta gives rise to the electromagnetic forces, for example, appears to have never been
carried out. Having a detailed description of the mechanism of interaction between particles in the Standard
Model would have made it possible to examine whether it can provide an explanation of the origin of inertia.
Indeed, if the electromagnetic force between two charges, for instance, can be fully understood in terms of the
recoils to which a charge is subjected when emitting and absorbing virtual photons, then one can expect that
inertia might be explained by the Standard Model.

How the recoils from virtual photons absorbed by an elementary charge cause its inertia can be demonstrated
by considering a free charge. The recoils it suffers due to the emission and absorption of the virtual photons of
its own (quantized) electromagnetic field cancel out precisely due to the spherical symmetry of the field.
However, taking into account the general-relativistic frequency shift of the incoming virtual photons that are
\textit{absorbed} by a \textit{non-inertial} charge (accelerated or supported in a gravitational field) implies that
the recoils imparted on the charge by the absorbed (``blue'' and ``red'' shifted) virtual photons of its own field
do not cancel out. That imbalance in the recoils gives rise to a self-force which is electromagnetic in origin and
which acts on the non-inertial charge \cite[Chap. 9]{petkov}. As a result the charge \textit{resists} its
acceleration or its being prevented from falling in a gravitational field. In other words, the non-inertial charge
resists its deviation from its geodesic path, which implies that the self-force manifest itself as an
electromagnetic component of the inertial force.

As according to the Standard Model the strong and weak interactions are likewise mediated by virtual quanta it
follows that a non-inertial elementary particle that is involved in weak and strong interactions will be also
subjected to the imbalanced recoils from the virtual quanta of its own strong and weak ``fields'', which are
absorbed by the non-inertial particle. This mechanism implies that all interactions in the framework of the
Standard Model contribute to the inertia and mass of elementary particles. That contribution might either
complement the Higgs mechanism or provide an alternative mechanism.

It should be emphasized that the \textit{same} mechanism that accounts for inertia and mass of the classical
electron -- the interaction of the electron charge with its own field -- appears to be at work at the quantum
level as well. This would explain why the electron theory correctly predicted the existence of electromagnetic
inertia and would demonstrate that if the electron theory had not been prematurely abandoned its careful
examination would have naturally led to the mechanism that gives rise to inertia and mass at the quantum level.

\textit{1.2. Worldtubes and anisotropic velocity of light in non-inertial reference frames.} The missed
opportunity to have a complete understanding of propagation of light in non-inertial reference frames is another
example of underestimation of the power of conceptual analysis and not taking seriously Minkowski's view that
physical objects are worldtubes in spacetime. Try to recall how many books on relativity define \textit{average
anisotropic velocity of light} in accelerating frames or in a gravitational field. Of course, the Sagnac effect and
the time delay (Shapiro) effect, for example, are nicely discussed but they have never been explained properly
in terms of the average anisotropic velocity of light.

Although deriving the expression for the average anisotropic velocity of light is straightforward \cite[Chap.
7]{petkov} realizing the need for such a velocity requires just a little trust in conceptual analyses -- the type of
analyses that made Galileo and Einstein, for instance, such great scientists.

\begin{figure}[h]
\centering
\includegraphics[height=5.5cm]{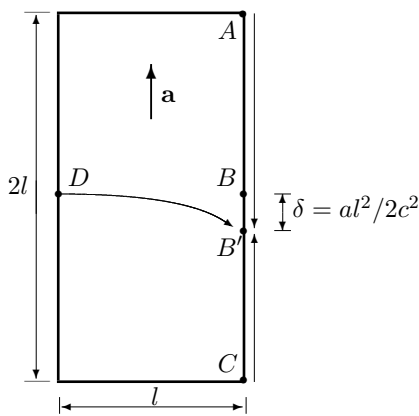} \caption{Propagation of light in an accelerating Einstein elevator.}\label{elev}
\end{figure}

This can be demonstrated by considering the famous example of Einstein's elevators -- accelerating or at rest in
a gravitational field. We will do it for the case of an accelerating elevator; the same analysis applies for an
elevator supported in a gravitational field. Fig. 2 depicts an elevator which accelerates with an acceleration $a$.
Two light rays emitted from $A$ and $C$ toward the middle point of the elevator $B$ are introduced (in
addition to the original horizontal light ray) in order to demonstrate that the average velocity of light in an
accelerating frame is not $c$ (but the local velocity is, of course, $c$). The three light rays are emitted
\textit{simultaneously\/} in the elevator from points $D$, $A$, and $C$ toward point $B$. Let $I$ be an inertial
reference frame instantaneously at rest with respect to the elevator (i.e., the instantaneously comoving frame)
at the moment the light rays are emitted. As $I$ and the non-inertial frame $N$ of the elevator share a
common instantaneous three-dimensional space and therefore common simultaneity at the moment the three
light signals are emitted, the emission of the rays is simultaneous in $N$ as well as in $I$. At the next moment
an observer in $I$ sees that the three light rays arrive simultaneously not at point $B$, but at $B^\prime$,
since for the time $t=l/c$ the light rays travel toward $B$, the elevator moves a distance $\delta
=at^2/2=al^2/2c^2$. As the simultaneous arrival of the three rays at point $B^\prime$ as viewed in $I$ is an
absolute fact due to its being a \emph{single\/} event, it follows that the rays arrive simultaneously at
$B^\prime$ as seen in $N$ as well. Since for the \emph{same\/} coordinate time $t=l/c$ in $N$, the three light
rays travel \emph{different\/} distances $DB^\prime\approx l$, $AB^\prime=l+\delta$, and
$CB^\prime=l-\delta $, before arriving simultaneously at point $B^\prime$, an observer in the elevator
concludes that the propagation of light is affected by the elevator's acceleration. As seen in Fig. 2 an observer
in $N$ concludes (to within terms $\sim c^{-2}$) that the \emph{average} velocities of the light rays
propagating from $A$ to $ B^{\prime }$ and from $C$ to $ B^{\prime }$ are slightly greater and smaller than
$c$, respectively.

\begin{figure}[h]
\centering
\includegraphics[height=4cm]{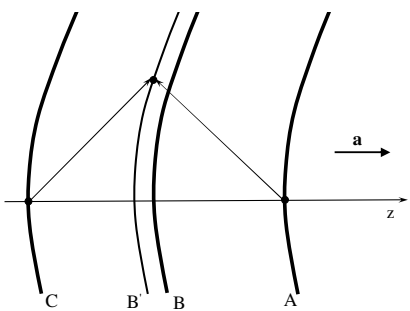} \caption{The simultaneous arrival of the vertical light rays at
point $B^{\prime }$ in Fig. 2 is self-evident when it is taken into account that the points
$A$, $B$, $C$, and $B^{\prime }$ are worldlines, which are \textit{curved}.}\label{alight}
\end{figure}

One can arrive at the idea that the average velocity of light in non-inertial frames is not $c$ also by taking into
account that the light rays and the points $A$, $B$, $C$, and $B^{\prime }$ are in reality worldlines in
spacetime. It is clearly seen in Fig. \ref{alight} that the light rays emitted from $A$ and $C$ arrive
simultaneously at $B^{\prime }$, not at $B$, due the fact that the worldlines of the points $A$, $B$, $C$, and
$B^{\prime }$ are \textit{curved}. When the curvature of the worldtubes of the emitter and the receiver in the
gravitational redshift (Pound--Rebka) experiment is taken into account one can easily overcome a widely shared
misconception -- that the redshift effect proves the curvature of spacetime \cite{misner}. What it proves is that
the worldtubes of the emitter and the receiver are curved (not geodesic).

Overlooking the introduction of an average velocity of light in non-inertial frames has led to another missed
opportunity -- a derivation of an anisotropic volume element, which is necessary for calculating electromagnetic
phenomena \textit{directly} in such frames. Using the anisotropic volume element in those calculations
simultaneously solved two old problems -- the appearance of (i) a factor of $1/2$ in Fermi's potential of a charge
in a gravitational field, and (ii) a factor of $4/3$ in the expression of the self-force acting on a non-inertial
classical electron \cite[Chap. 8]{petkov}.

\textit{1.3. Worldtubes and the discovery of general relativity.} This section is specifically devoted to the young
generation of physicists. It is sometimes claimed that future researchers cannot be taught how to make
discoveries since discoveries are logical jumps coming as sudden insights to individual researchers. Often
general relativity is given as an excellent example. Moreover, Einstein himself described the realization that a
person falling from the roof of a house does not feel the force of gravity as the happiest thought in his life.
Such insights do exist in the history of science but not all advances in science follow that pattern -- it is
sufficient to mention how conceptual analyses helped Galileo to arrive at the idea of inertial motion (by
disproving Aristotle's view of motion) and at his principle of relativity. Even Einstein himself convincingly proved
that conceptual analyses are physics at its best by his thought experiments.

General relativity is indeed an excellent example but not of the view that discoveries are results of happy
insights; it is an example of making discoveries by having a productive view on the nature of physical theories
and by employing a rigorous conceptual analysis of the available theoretical and experimental evidence.

Minkowski's representation of special relativity makes it possible to arrive at the idea of gravity as a
manifestation of the non-Euclidean geometry of spacetime not just naturally, but inescapably. A conceptual
analysis of Newton's gravitational theory could and should have revealed, long before Einstein realized it, that a
falling body offers no resistance to its acceleration. This means that the body is not subjected to any
gravitational force, which would be necessary if the body resisted its fall. Therefore, the falling body moves
non-resistantly, by inertia. But how could that be since it accelerates?

Taking seriously the existence of worldtubes and its implication that inertia originates from a four-dimensional
stress that arises in the deformed worldtube of a non-inertial body provide a radical resolution -- the worldtube
of the falling body should be curved (reflecting the fact that the body accelerates), but not deformed
(accounting for the fact that the body does not resist its acceleration). Such a worldtube can exist only in a
curved spacetime where the geodesic worldtubes of bodies moving by inertia (non-resistantly) are
\textit{curved but not deformed}.

\textbf{2. The major missed opportunity in quantum physics.} No one knows what the quantum object is. What
is worse is that the standard interpretation of quantum mechanics tells us that we cannot say or even ask
anything about the quantum object between measurements. In this sense, I think, Einstein was right that
quantum mechanics is essentially incomplete. It is unrealistic to assume that an electron, for example, does not
exist between measurements. But if it exists, it is something and we should know what that something is. In
general, we should try to understand what the (free or interacting) quantum object is.

Although relativity does not fully apply at the quantum level since its equations of motion manifestly fail to
describe the behaviour of quantum objects, spacetime does seem to be the arena of the quantum world as
well. Then the question is: ``Are quantum objects also worldlines in spacetime?'' It is well known that the
answer is negative.

This can be demonstrated in the case of interference experiments performed with single electrons
\cite{tonomura}. In such double--slit experiments accumulation of successive single electron hits on the screen
builds up the interference pattern that demonstrates the wave behaviour of \textit{single} electrons. When
looking at the screen, every single electron is detected as a localized entity and the natural question is whether
the electron was such an entity before it hit the screen. Our intuition leads us to assume that if the electron hits
the screen as a localized entity, it is such an entity at \textit{every} moment of time, which implies that the
electron exists \textit{continuously} in time as a localized entity. But if this were the case, every single electron
would behave as an ordinary particle and should go only through one slit and no interference pattern would be
observed on the screen. Therefore, the electron cannot be a localized entity at all moments of time, i.e. it
cannot be a worldline in spacetime.

If this had been realized (and it could have been realized if more trust had been put in conceptual analyses), it
could not have been so difficult to ask whether an electron exists \textit{continuously} in time if it is not a
worldline. Such a question could have provided a new perspective at looking at quantum phenomena and
apparent paradoxes.

\textbf{3. Possible missed opportunities to test string theory.} I think string theory is perhaps the best example
of overestimation of the predictive power of mathematics in physics. String theory constitutes an unprecedent
case in the history of physics -- while string theorists all admit that the theory has not yet been experimentally
confirmed they behave as if this has already been done (or will certainly be done) and string theory has been
treated on equal footing with the accepted physical theories.

A necessary condition that should be met by any newly proposed physical theory is not contradicting the existing
reliable theoretical and experimental evidence. It is precisely here where I think opportunities to test string
theory might have been missed. Let me give just one example. While string theory has extensively studied how
the interactions in the hydrogen atom can be represented in terms of the string formalism, I wonder how string
theory would answer a much simpler question -- what should be the electron in the ground state of the
hydrogen atom in order that the hydrogen atom does not possess a dipole moment in that state?

\end{document}